\documentclass[12pt]{article}
\usepackage{epsfig,amsfonts,amssymb}
\usepackage{hyperref}
\usepackage{cite}
\input epsf.sty
\usepackage{cite}

\def\ZZZ{{\hbox{ Z\kern-1.6mm Z}}}
\def\RRR{{\hbox{ R\kern-2.4mm R}}}
\def\CCC{{\hbox{ C\kern-2.0mm C}}}
\def\zzz{{\hbox{z\kern-1mm z}}}

\newcommand{\qeq}{{\hbox{=\kern-2.3mm ? \kern.5mm }}}
\renewcommand{\qeq}{=}

\newcommand{\wh}{\widehat}

\newcommand{\NN}{{\cal N}}

\newcommand{\be}{\begin{equation}}
\newcommand{\ee}{\end{equation}}
\newcommand{\ben}{\begin{eqnarray}\displaystyle}
\newcommand{\een}{\end{eqnarray}}

\newcommand{\bea}[1]{\begin{eqnarray}\label{#1} }
\newcommand{\eea}{\end{eqnarray}}

\newcommand{\refb}[1]{(\ref{#1})}

\def\one{{\hbox{ 1\kern-.8mm l}}}
\def\zero{{\hbox{ 0\kern-1.5mm 0}}}

\begin{document}

\baselineskip 24pt

\begin{center}
{\Large \bf 
Wall Crossing Formula for $\NN=4$ Dyons:

A Macroscopic Derivation}

\end{center}

\vskip .6cm
\medskip

\vspace*{4.0ex}

\baselineskip=18pt

\centerline{\large \rm   Ashoke Sen }

\vspace*{4.0ex}

\centerline{\large \it Harish-Chandra Research Institute}

\centerline{\large \it  Chhatnag Road, Jhusi,
Allahabad 211019, INDIA}

\vspace*{1.0ex}
\centerline{E-mail:  sen@mri.ernet.in, ashokesen1999@gmail.com}

\vspace*{5.0ex}

\centerline{\bf Abstract} \bigskip

We derive the wall crossing formula for the decay of
a quarter BPS dyon into a pair of half-BPS dyons  
by analyzing the quantum
dynamics of multi-centered black holes
in
$\NN=4$ supersymmetric string theories. 
Our analysis encompasses 
the cases where
the final decay products are non-primitive dyons. The results are
in agreement 
with the  microscopic formula for the dyon spectrum
in the special case
of heterotic string theory on $T^6$.

\vfill \eject

\baselineskip=18pt

Much of the study of string theoretic black holes has focussed on
BPS black holes since the degeneracy -- or more precisely an
appropriate index -- associated with BPS states are protected and
do not change under a continuous variation of the various moduli.
Nevertheless it has been known for some 
time\cite{9407087,9408099,9712211} 
that in some cases
the index can jump across codimension one subspaces of the moduli
space on which the original state becomes marginally unstable
against possible decay into a pair of BPS states.
Such  subspaces are known as walls of marginal stability, and the
jump in the index across these walls -- known as the wall crossing
formula -- has been the subject of intense investigation in recent
years\cite{0005049,0010222,0101135,0206072,
0304094,0702146,0706.3193,kont,private}. 
While most of the work has been focussed on half-BPS dyonic
black holes
in $\NN=2$ supersymmetric string theories, by now we also have
a good understanding of this phenomenon in $\NN=4$ and
$\NN=8$ supersymmetric string 
theories\cite{0605210,0702141,0702150,0705.3874,
0706.2363,0802.0544,0802.1556,0803.1014}. 
In fact in the latter
theories the situation is somewhat better since we also have a
good understanding of the exact spectrum of BPS dyons in these
theories. This allows us to verify the general wall crossing formula
derived from macroscopic considerations involving multi-centered
black hole solutions\cite{0206072}
against explicit results obtained from microstate 
counting\cite{0705.3874,0706.2363}.

For deriving the wall
crossing formula for walls on which the decay products
carry primitive charge vectors, 
the original techniques of \cite{0206072} involving analysis of 
multi-centered
black hole solutions are
equally applicable to theories with $\NN=2$, 4 and 8 
supersymmetries in four dimensions. In the case of $\NN=2$
supersymmetric theories these as well
as other techniques have been developed 
which allow us to generalize the wall crossing formula to the
cases where the final decay products carry non-primitive charge
vectors\cite{0702146,0706.3193,kont}. 
However so far these techniques have not been generalized
to $\NN>2$ supersymmetric string theories.

Recently by examining the exact formula for the dyon partition
function
in the special case of heterotic string theory on $T^6$, 
Refs.\cite{0802.0544,0802.1556} studied 
the jump in the dyon spectrum of this
theory across various walls of marginal stability, including the
ones on which the decay products are non-primitive, and proposed
a general wall crossing formula for such 
walls. The purpose of this
paper is to give a macroscopic derivation of this formula from
the study of multi-centered black holes in a general $\NN=4$
supersymmetric string theory. As we shall see, this is possible and
yields results in perfect agreement with the results derived from
microscopic analysis.\footnote{The formula 
for the dyon partition function was originally guessed in
\cite{0802.0544,0802.1556} by imposing various consistency
conditions including known wall crossing formula for
decay into  primitive dyons. 
More recently a proof has been suggested in \cite{0803.2692}.} 
Since the jump in the index across these
walls is exponentially small compared to the leading contribution,
this reinforces our belief  that black holes capture not only the
leading contribution to the statistical
entropy but also the exponentially
suppressed contributions.

We begin by reviewing the
derivation of the wall crossing formula in 
$\NN=2$ supersymmetric string theories. 
Let us consider the wall of marginal
stability associated with the 
decay of a dyon of charge $(Q,P)$
into a pair of
primitive dyons of charges $(Q_1,P_1)$ and $(Q_2,P_2)$, 
with $Q$ and $P$ denoting electric and magnetic charges in some
basis.  In this case the jump in the index across this
 wall of marginal
stability can be computed from the following simple 
argument\cite{0005049,0010222,0101135,0206072,0304094,0702146}.
A classical analysis shows that a two centered solution of total
charge $(Q,P)$, with one center having charges $(Q_1,P_1)$ and the
other center having charges $(Q_2,P_2)$, exist on one side of the
wall and does not exist on the other side\cite{0005049,0010222}. 
Thus the jump in the index
can be identified as the index associated with the two centered
solution. One also finds that as we approach the wall, the 
separation between the two centers approaches infinity. Thus in this
limit the index can be identified as the product of the index associated
with each component, and the index associated with the supersymmetric
quantum mechanics describing relative motion between the two
centers. The latter gives a contribution of
$(-1)^{Q_1\cdot P_2-Q_2\cdot P_1+1} \, |Q_1\cdot P_2-Q_2\cdot P_1|$.
Thus if $d_h(Q_1,P_1)$ and $d_h(Q_2,P_2)$ denote the index 
associated with the decay products, the net change in the index
will be given by
\be \label{ei1}
(-1)^{Q_1\cdot P_2-Q_2\cdot P_1+1} \, |Q_1\cdot P_2-Q_2\cdot P_1|\,
d_h(Q_1,P_1)\, d_h(Q_2,P_2)\, .
\ee
When the charge vectors $(Q_1,P_1)$ and/or $(Q_2,P_2)$ are
non-primitive, the above formula is known to undergo non-trivial
modification\cite{0702146,0706.3193,kont}.

Our goal in this paper will be to follow a similar logic for deriving the
change in the index across a wall of marginal stability where a quarter
BPS state in an $\NN=4$ supersymmetric string theory decays into a
pair of half-BPS states. As we shall see, this procedure can give the
wall crossing formula non only for decay into a pair of primitive
dyons but also for decay into a pair of non-primitive dyons. 
We begin by introducing some notations. For a given charge vector
$(Q,P)$ we define helicity trace $B_{n}$ via the 
relation\cite{9708062,9708130}
\be \label{ei2}
B_{n}(Q,P) = {1\over n!}
Tr_{(Q,P)}\left( (-1)^F (2h)^{n}\right)\, ,
\ee
where $Tr_{(Q,P)}$ denotes trace over states with charge
$(Q,P)$, $F$ denotes the fermion number and 
$h$ denotes the helicity
of the state. Since a quarter BPS state breaks 12 out 
of the 16 supersymmetries, it has 12 fermion zero modes and we need
at least 6 factors of $2h$ to get a non-zero answer
for the supertrace. We shall denote  by $d(Q,P)$ the index
$-B_6(Q,P)$ associated with quarter BPS states. A typical quarter
BPS supermultiplet contains states of helicities between 
$H-{3\over 2}$ and $H+{3\over 2}$, with the state of helicity 
$H+{s-3\over 2}$ coming with a degeneracy of ${6\choose s}$.
It is easy to see that the contribution of these states to $B_6$ is
given by $(-1)^{2H+1}$, with 
the $1/6!$ in the normalization factor of
$B_6$ cancelling a factor of $6!$ coming from
sum over $s$.
Thus
$d(Q,P)$ effectively counts the number of quarter BPS
supermultiplets
carrying charge $(Q,P)$ weighted by $(-1)^{2H}$, $H$ being the
average helicity of all the states in the supermultiplet.
On the
other hand half-BPS states break 8 out of the 16 supersymmetries,
and hence we need 4 factors of $2h$ to get a non-vanishing 
supertrace. We denote by $d_h(Q_i,P_i)$ the index 
$B_4(Q_i,P_i)$
associated with the half-BPS decay products. $d_h(Q_i,P_i)$
counts the number of half-BPS supermultiplets weighted
by $(-1)^{2H}$, $H$ being the average helicity of all the states
in the half-BPS supermultiplet. Our goal is to calculate the
change in $d(Q,P)$ across a wall of marginal stability in terms
of $d_h(Q_i,P_i)$ associated with the half-BPS decay products.

First consider the case when both decay products are 
primitive.\footnote{I wish to thank F.~Denef for 
discussion on this case.}
In this case 
on one side of the wall of marginal stability we have a two centered
classical black hole solution, with one center carrying charge
$(Q_1,P_1)$ and the other center carrying charge 
$(Q_2,P_2)$.\footnote{Since both the final dyons are half-BPS
states of the full $\NN=4$ supersymmetry algebra, they are
small black holes\cite{9504147,0409148}, 
but this does not affect our argument.} 
This solution ceases to exist on the other side of the wall; hence the
jump in $d(Q,P)$ across this wall can be identified as the
contribution to $-B_6(Q,P)$ from this two centered solution.
Now since the wall crossing formula is a change in the index, we expect
it to be invariant under a continuous change in the 
moduli.\footnote{In $\NN=2$ supersymmetric string theories the
decay products are half-BPS and can themselves decay across walls
of marginal stability. As a result the wall crossing formula also
changes as a function of the moduli. In contrast in the $\NN=4$
supersymmetric string theories the decay products are half-BPS
states whose index does not change as we vary the moduli. Thus
we expect the wall crossing formula to be unchanged as we vary the
moduli. \label{ff1}}
This allows us to
work in a
region of the moduli space where one of the decay products
(say with charge $(Q_1,P_1)$)
is heavy and the other (with charge $(Q_2,P_2)$)
is light. In this case 
we can describe the system as the light particle moving in the
background of the heavy particle, and ignore the backreaction
of the light particle on the dynamics of the heavy particle. 
We now note that since the
heavy particle breaks 8 out of 16 supersymmetries, in order to
get a non-vanishing contribution to the
supertrace over the states of the heavy particle we must insert a
factor of $(2h_{(1)})^4$ into the trace, 
$h_{(1)}$ being the helicity of the heavy particle. 
Since the
light particle moves in the background produced by the heavy
particle, it
only feels 8 of the unbroken supersymmetries. Furthermore since
the classical two centered solution is quarter BPS, the light particle
breaks 4 out of these 8 supersymmetries. As a result it carries 4
fermion zero modes, and we must insert a factor of $(2h_{(2)})^2$
into the supertrace over the light particle degrees of freedom to get
a non-vanishing answer. On the other hand since in flat space-time
the light particle, being half-BPS, 
breaks 8 out of 16 supersymmetries, it has altogether 8 fermion
zero modes in flat space-time. 4 of these zero modes must be lifted,
\i.e.\ take part in the interaction, in the presence of the heavy
particle. In fact these
must combine with the bosonic modes describing  the physical
coordinates of the light particle
to describe a supersymmetric
quantum mechanics with 4 supersymmetries since the final
configuration has four unbroken
supersymmetries. If $d_{rel}$ denotes the 
number of supersymmetric
ground states of this quantum mechanical system, weighted by
$(-1)^F$, then the index $-B_{6}$ of the two centered
dyon system will be given by
\be \label{ei3}
d_h(Q_1,P_1) \, d_h(Q_2, P_2)\, d_{rel}\, .
\ee
The normalization and sign 
factors work out as follows. First of all we
note that the
coefficient of $(2h_{(1)})^4 (2h_{(2)})^2$ term in 
$(2h_{(1)}+2h_{(2)})^6$
is ${6\choose 4}$. This 
factor of ${6\choose 4}$ combines with the
$1/6!$ in the definition of $B_6$
to give $1/4!2!$. The $4!$ now cancels the trace of 
$(-1)^F (2h_{(1)})^4$
over the 8
fermion zero modes carried by the heavy state and $2!$ cancels
the trace of $(-1)^F (2h_{(2)})^2$ 
over the fermion zero modes carried by the light state, leaving
behind a minus sign. This minus sign compensates for the
minus sign in the relation between $d(Q,P)$ and $B_6(Q,P)$.

In order to calculate $d_{rel}$ we can
work in a subspace of the moduli space where the original dyon
with charge $(Q,P)$ and the decay products
carrying charges $(Q_1,P_1)$ and $(Q_2,P_2)$ can be regarded
as half-BPS dyons of an $\NN=2$ subalgebra. In this case
we can 
examine the situation from the point of
view of $\NN=2$ supersymmetry.  
In the absence of the heavy particle
the light particle would break 4 of the 8 supersymmetries and hence
would carry four fermion zero modes. However in the presence of the
heavy particle these fermion zero modes will be lifted since the
heavy particle already breaks 4 of the 8 supersymmetries and hence
the light particle will not break any further supersymmetry. Thus
effectively the 4 fermion zero modes of the light particle will become
interacting and combine with the bosonic coordinates to give a
supersymmetric quantum mechanics with four supersymmetries. This
must be the same interacting quantum mechanical system that 
we got by analyzing
the system from the $\NN=4$ viewpoint, -- the effect of truncation to the
$\NN=2$ subsector being simply 
the removal of the 4 non-interacting fermion
zero modes on the light state. Now we can use the already existing results
in $\NN=2$ theory\cite{0206072} 
to conclude that this supersymmetric quantum
mechnics has a
set of supersymmetric ground states of Witten index 
$(-1)^{Q_1\cdot P_2-Q_2\cdot P_1+1} \, |Q_1\cdot P_2-Q_2\cdot P_1|$.
Using \refb{ei3} we now get the total
jump in the index across the wall
\be \label{ei4}
\Delta d(Q,P) = 
(-1)^{Q_1\cdot P_2-Q_2\cdot P_1+1} \, |Q_1\cdot P_2-Q_2\cdot P_1|
\, d_h(Q_1,P_1)\, d_h(Q_2,P_2)\, .
\ee
Note that the final formula is symmetric under the exchange of final
state charges even though we treated them differently in our analysis.

This finishes our analysis of the wall crossing formula for decay into
a pair of primitive dyons. Now consider the case where the light
dyon carries a non-primitive charge vector, \i.e.\ $(Q_2,P_2)$
is given by some integer $N_2$ times a a primitive charge vector
$(Q_2/N_2, P_2/N_2)$. In this case besides the two centered
configuration described earlier, the system also admits 
multi-centered
configurations, with the first center carrying charges $(Q_1,P_1)$
and the others carrying charges $(\alpha_i Q_2/N_2, 
\alpha_i P_2/N_2)$,
with $\alpha_i\in\ZZZ$, $\alpha_i\ge  1$,
$\sum_i{\alpha_i}=N_2$. All of these configurations cease to exist on
the other side of the wall and hence could contribute to the
wall crossing formula.
Since we are working 
in a region of the moduli space where  the dyons
carrying charges of the form $(\alpha_i Q_2/N_2,
\alpha_i P_2/N_2)$ are light, they are also weakly interacting.
Thus we can regard the states of the full system as
tensor products of the states of the
heavy particle of charge $(Q_1,P_1)$ 
and the states of light
particles carrying charges $(\alpha_i Q_2/N_2, 
\alpha_i P_2/N_2)$  moving in the
background of the heavy particle. However the correponding trace
will vanish\cite{private} 
since the heavy particle carries 8 fermion zero modes
requiring insertion of 4 powers of helicity into the supertrace
and each of the light particles carries 4 fermion zero modes, requiring
insertion of 2 powers of helicity into trace over the Hilbert space
of each particle. 
Since we only have a total of 6 powers
of helicity in the definition of $B_6$ we cannot saturate all the
fermion zero modes..

There is however an exception to this rule.
The above argument
assumes that the full 
Hilbert space is a direct product of the Hilbert spaces of 
the component
dyons.
However if some of the components are
identical then we must (anti-)symmetrize the wave-function and the
full Hilbert space is no longer a direct product of the
Hilbert spaces of the component dyons. 
Such configurations could give non-vanishing
contribution to the index.\footnote{Similar
issues arose in the analysis of \cite{0803.2692}.}
We begin by examining the contribution from a
multi-centered dyon configurations with one
heavy center of charge $(Q_1,P_1)$ and $L$
light centers each carrying
charges $(Q_2/L, P_2/L)$, $L$ being a factor
of $N_2$. Furthermore we take all these $L$ light
centers in the same internal quantum state,\footnote{By
internal quantum state of a half-BPS dyon we shall refer to
supersymmetry singlet
part of the state. We tensor this state with the states obtained by
quantizing the fermion zero modes to get the full supermultiplet.}
and in the
same state of the supersymmetric quantum mechanics describing
the motion of the light particle in the heavy particle background.
We shall argue later that these are the only types of multi-centered
configurations which contribute to the index.

Let us now compute the contribution to the index from 
such a configuration. We denote by $h_{int}$ the contribution
to the helicity of one of the $L$ light
dyons  from the internal quantum state.
This dyon also gets
a contribution to the helicity  from the supersymmetric quantum
mechanics describing
the motion of the light particle in the heavy particle background,
-- this is given by\cite{0206072}
\be \label{ef0}
h_{rel}=
|Q_1\cdot P_2-Q_2\cdot P_1|/(2L) - (1/2)-{\rm integer}\, .
\ee
Finally quantization of the 4 fermion zero modes produces a 
4-fold degenerate state, with 1 state of helicity $-1/2$, two states
of helicity 0 and one state of helicity $1/2$. Let us denote 
these four states by $|i\rangle$ with $1\le i\le 4$
and let $\hat h_i$ be the contribution to the
helicity of
the $i$th state from the fermion zero modes. Thus when we
take the tensor product of a fixed
internal state of the light dyon, a fixed
state of the supersymmetric quantum mechanics, and the
states obtained by quantizing the fermion zero mode, the
states
can be labelled by the index $i$, with total helicity
\be \label{ef1}
h_i= h_{int} + h_{rel} + \hat h_i\, .
\ee
We shall refer to these  states as single particle states.
Our goal is to consider an $L$ particle state, with each of the $L$
particles being in the same internal state and same state of the
supersymmetric quantum mechanics, and calculate the contribution
to $B_6$ from these states. For this we need to first
identify the
statistics of the single particle states described above. Naively
the particle will be bosonic or fermionic depending on whether
$h_i$ is integer or half integer. However 
in the contribution to $h_{rel}$ given in \refb{ef0}, the
part proportional to $|Q_1\cdot P_2-Q_2\cdot P_1|/(2L)$
comes from the angular momentum of the 
electromagnetic field of the dyon system and does not directly
contribute to the statistics of the light particle. 
Thus the particle should be regarded
as bosonic or fermionic depending on whether
$2h_{int}+2\hat h_i+1$ is even or odd. The $L$ particle
states are then labelled by specifying the occupation number
$n_i$ of the $i$th state, subject to the conditions
that $\sum_i{n_i}=L$,  $n_i$ takes values 0 and 1 if
$2h_{int}+2\hat h_i+1$ is odd 
and $n_i$ takes all non-negative integer values
if $2h_{int}+2\hat h_i+1$ is even. 

Let us now denote by
$g_L$ the quantity:
\be \label{ei5}
g_L 
= -{1\over 2!} Tr_{L} \left( (-1)^{2h}\, (2h)^2\right)\, ,
\ee
where $Tr_L$ denotes trace over the Hilbert space of $L$
particle states introduced above and $h$ in \refb{ei5} stands
for the total contribution to the helicity from the $L$ light
dyons.
Then the contribution to $-B_6$ from these
states will be given by the product of $g_{L}$ 
and $d_h(Q_1,P_1)$.
{}Using the description of the multiparticle
states given above, we get
\be \label{ei6}
g_L = -{1\over 2}
\sum_{\{n_i\}, \sum_i{n_i}=L}\, 
(-1)^{2\sum_i{n_i h_i}} \, \left(2\sum_k n_k h_k\right)^2
\, .
\ee
Using \refb{ef1} we can write this as
\be \label{ef3}
g_L = -(-1)^{(2h_{rel}-1) L} \, \wh g_L\, ,
\ee
where
\be \label{ef4}
\wh g_L = {1\over 2}\,
\sum_{\{n_i\}, \sum_i{n_i}=L}\, 
(-1)^{\sum_i{n_i (2h_{int}+2\hat h_i +1)}} \, \left(2\sum_k n_k (h_{int}
+ \hat h_k + h_{rel})\right)^2
\, .
\ee
$\wh 
g_L$ is most conveniently evaluated by first calculating the partition
function
\be \label{ei7}
f(\mu,\beta) \equiv \sum_{\{n_i\}} \, 
(-1)^{\sum_i{n_i (2h_{int}+2\hat h_i+1)}} \, e^{2\beta\sum_i n_i 
(h_{int}+\hat h_i+ h_{rel})}
\, e^{\mu \sum_i n_i}\, ,
\ee
and then calculating $\wh g_L$ as the coefficient of the $e^{\mu L}$ term
in
\be \label{ei8}
{1\over 2}\, 
\left[{d^2\over d\beta^2} f(\mu,\beta)\right]_{\beta=0}\, .
\ee
As described earlier, the sum over $n_i$ is restricted to 0 and 1
for $2h_{int}+2\hat h_i+1$ odd and to all non-negative integers
for $2h_{int}+2\hat h_i+1$ even. This gives
\be \label{eff1}
f(\mu,\beta) = \prod_i \left( 1 -
e^{\mu + 2\beta (h_{int}+\hat h_i + h_{rel})}
\right)^{(-1)^{2h_{int}+2\hat h_i}}\, .
\ee
Thus
\ben \label{eff2}
\ln f(\mu,\beta) 
&=& \sum_i \, (-1)^{2h_{int}+2\hat h_i}\, 
\ln \left( 1 -
e^{\mu + 2\beta (h_{int}+\hat h_i + h_{rel})}
\right) \nonumber \\
&=& -\sum_{k=1}^\infty \, {1\over k}\, 
\sum_i \, (-1)^{2h_{int}+2\hat h_i}\,
e^{k \left(\mu + 2\beta (h_{int}+\hat h_i + h_{rel})\right)}
\nonumber \\
&=& \sum_{k=1}^\infty \, 
{1\over k}\, (-1)^{2h_{int}} \, e^{k \left(\mu + 2\beta (h_{int} 
+ h_{rel})\right)}\, (e^{k\beta/2} - e^{-k\beta/2})^2\, ,
\een
where in the second step we have expanded $\ln(1-x)$ in a
Taylor series expansion in $x$ and in the
last step we have explicitly carried out the sum over $i$
using the fact that there is one state with $\hat h_i=-1/2$, two states
with $\hat h_i=0$ and one state with $\hat h_i=1/2$. This gives,
from \refb{ei8}
\be \label{egg1}
\hat g_L = (-1)^{2 h_{int}}\, L\, ,
\ee 
and hence, from \refb{ef0}, \refb{ef3},
\be \label{egg2}
g_L = (-1)^{Q_1\cdot P_2-Q_2\cdot P_1+2 h_{int}+1}\, L\, .
\ee
As already mentioned below \refb{ei5}, 
the net contribution to the index $-B_6$
from the specific configurations analyzed above will be given by
$g_L\, d_h(Q_1,P_1)$.

The total contribution to $-B_6$ from the configurations
where the light dyon is split into $L$ identical centers is
obtained by summing $g_L\, d_h(Q_1,P_1)$ 
over all the internal states of the dyon
of charge $(Q_2/L, P_2/L)$ and all the
supersymmetric ground states of the supersymmetric 
quantum mechanics describing the motion of a single particle of
charge $(Q_2/L, P_2/L)$ in the background of the heavy particle
with charge $(Q_1,P_1)$.
The former gives a factor of $d_h(Q_2/L, P_2/L)$ 
while the latter 
gives a factor of $|Q_1\cdot P_2-Q_2\cdot P_1|/L$\cite{0206072}.
This gives the net contribution to the index from these configurations 
to be
\be \label{egg3}
(-1)^{Q_1\cdot P_2-Q_2\cdot P_1+1}\, 
|Q_1\cdot P_2-Q_2\cdot P_1|\, d_h(Q_1, P_1)
d_h(Q_2/L, P_2/L)\, .
\ee
Note that $(-1)^{2h_{int}}$ has been absorbed into the definition
of $d_h(Q_2/L, P_2/L)$.
Finally we must sum over all possible values of $L$ since all
multi-centered configurations with one center having charge $(Q_1,P_1)$
and the other $L$ centers having charges $(Q_2/L, P_2/L)$ 
will disappear
across the wall of marginal stability on which the original dyon
decays into dyons of charge $(Q_1,P_1)$ and $(Q_2,P_2)$. This
gives the final formula for the jump in the index to be
\be \label{egg4}
(-1)^{Q_1\cdot P_2-Q_2\cdot P_1+1}\, 
|Q_1\cdot P_2-Q_2\cdot P_1|\, \sum_{L|(Q_2,P_2)} \,
d_h(Q_1, P_1)
d_h(Q_2/L, P_2/L)\, .
\ee

One could in principle consider more general configurations
where the charge vector $(Q_2,P_2)$ splits into different groups
with the members within each group being identical 
but members of
different groups being not identical. In this case each group will
require an insertion of an $h^2$ factor to saturate its fermion 
zero modes. Since a factor of $h^4$ is already used up by the 
heavy state we see that we can allow at most one group. Thus
the configurations 
we have analyzed above are the only ones which can
contribute to $B_6$.

This finishes our analysis of the wall crossing formula when one
of the decay products is non-primitive. What about the case
when both decay products are non-primitive? For this we
let both
$(Q_1,P_1)$ and $(Q_2,P_2)$ be non-primitive and continue
to work in the corner of the moduli space where the dyon of charge
$(Q_1,P_1)$ is much heavier than the dyon of charge $(Q_2,P_2)$.
Now the dyon of charge $(Q_1,P_1)$ can also split into multiple
centers producing a non-spherically symmetric background for the
dyons of charge $(Q_2/L, P_2/L)$. However by arguments similar
to the ones given above we can conclude that unless all the centers
into which $(Q_1,P_1)$ splits are in the same quantum state, the
contribution to the index from these configurations will vanish.
If we use position space basis for these centers -- which is the natural
basis for heavy particles -- we see that the different centers into which
the dyon of charge $(Q_1,P_1)$ splits must coincide in space.
Thus it continues to produce a spherically symmetric potential
for dyons of charge $(Q_2/L,P_2/L)$  and our previous analysis
goes through except for a possible change in the overall
factor associated with the index of the dyon of charge $(Q_1,P_1)$.
Thus the jump in the index must take the form:
\be \label{es1}
(-1)^{Q_1\cdot P_2-Q_2\cdot P_1+1}\, 
|Q_1\cdot P_2-Q_2\cdot P_1|\, \sum_{L|(Q_2,P_2)} \,
f_h(Q_1, P_1)
d_h(Q_2/L, P_2/L)\, ,
\ee
for some function $f_h(Q_1,P_1)$. 
$f_h(Q_1, P_1)$ gets a contribution of $d_h(Q_1,P_1)$ from 
the single
dyon state but also possible additional contributions from the
multi-dyon states into which the dyon of charge $(Q_1,P_1)$ 
may split.
To determine the form of the
function $f_h$ we can go to the corner of the moduli
space where the dyon of charge $(Q_2,P_2)$ becomes heavy and
the dyon of charge $(Q_1,P_1)$ becomes light and repeat our
analysis. This gives the final form of the jump in the index
for a general decay where both $(Q_1,P_1)$ and $(Q_2,P_2)$
are non-primitive:
\be \label{es2}
\Delta d(Q,P)
= (-1)^{Q_1\cdot P_2-Q_2\cdot P_1+1}\, 
|Q_1\cdot P_2-Q_2\cdot P_1|\, 
\sum_{L_1|(Q_1,P_1)} d_h(Q_1/L_1, P_1/L_1)
\sum_{L_2|(Q_2,P_2)} \,
d_h(Q_2/L_2, P_2/L_2)\, .
\ee
This agrees with the result of \cite{0802.0544,0802.1556} 
derived from the microscopic
formula for the dyon partition 
function\cite{0802.0544,0802.1556,0803.2692} for the special
case of heterotic string theory
compactified on $T^6$.

\medskip

\noindent {\bf Acknowledgment:} We would like to thank
Nabamita Banerjee, Shamik Banerjee,
Justin David, 
Frederik Denef, Dileep Jatkar and Yogesh Srivastava
for useful discussions.



\begin{thebibliography}{99}



\bibitem{9407087}
  N.~Seiberg and E.~Witten,
  ``Electric - magnetic duality, 
monopole condensation, and confinement in N=2
  supersymmetric Yang-Mills theory,''
  Nucl.\ Phys.\  B {\bf 426}, 19 (1994)
  [Erratum-ibid.\  B {\bf 430}, 485 (1994)]
  [arXiv:hep-th/9407087].

\bibitem{9408099}
  N.~Seiberg and E.~Witten,
  ``Monopoles, duality and chiral 
symmetry breaking in N=2 supersymmetric
  QCD,''
  Nucl.\ Phys.\  B {\bf 431}, 484 (1994)
  [arXiv:hep-th/9408099].

\bibitem{9712211}
  O.~Bergman,
  ``Three-pronged strings and 1/4 BPS states in N = 4 super-Yang-Mills
  theory,''
  Nucl.\ Phys.\  B {\bf 525}, 104 (1998)
  [arXiv:hep-th/9712211].

\bibitem{0005049}
  F.~Denef,
  ``Supergravity flows and D-brane stability,''
  JHEP {\bf 0008}, 050 (2000)
  [arXiv:hep-th/0005049].

\bibitem{0010222}
F.~Denef,
``On the correspondence between D-branes 
and stationary supergravity solutions of type
II Calabi-Yau compactifications'', 
arXiv:hep-th/0010222.

\bibitem{0101135}
  F.~Denef, B.~R.~Greene and M.~Raugas,
  ``Split attractor flows and the spectrum 
of BPS D-branes on the 
quintic,''
  JHEP {\bf 0105}, 012 (2001)
  [arXiv:hep-th/0101135].
  
\bibitem{0206072}
  F.~Denef,
  ``Quantum quivers and Hall/hole halos,''
  JHEP {\bf 0210}, 023 (2002)
  [arXiv:hep-th/0206072].

\bibitem{0304094}
  B.~Bates and F.~Denef,
  ``Exact solutions for supersymmetric stationary black hole 
composites,''
  arXiv:hep-th/0304094.

\bibitem{0702146}
  F.~Denef and G.~W.~Moore,
  ``Split states, entropy enigmas, holes and halos,''
  arXiv:hep-th/0702146.

\bibitem{0706.3193}
  E.~Diaconescu and G.~W.~Moore,
  ``Crossing the Wall: Branes vs. Bundles,''
  arXiv:0706.3193 [hep-th].

\bibitem{kont}
M.~Kontsevich and Y.~Soibelman, talk given 
by M.~Kontsevich at the 
XXXVIIth Paris Summer Institute ``Black Holes, Black Rings and
    Modular Forms", August 13 to 24, 2007.

\bibitem{private}
F.~Denef, private communication.

\bibitem{0605210}
  J.~R.~David and A.~Sen,
  ``CHL dyons and statistical entropy function from D1-D5 system,''
  JHEP {\bf 0611}, 072 (2006)
  [arXiv:hep-th/0605210].

\bibitem{0702141}
  A.~Sen,
  ``Walls of marginal stability and dyon spectrum in N = 4 supersymmetric
  string theories,''
  arXiv:hep-th/0702141.

\bibitem{0702150}
  A.~Dabholkar, D.~Gaiotto and S.~Nampuri,
  ``Comments on the spectrum of CHL dyons,''
  arXiv:hep-th/0702150.
  

\bibitem{0705.3874}
  A.~Sen,
  ``Two Centered Black Holes and N=4 Dyon Spectrum,''
  arXiv:0705.3874 [hep-th].
    
\bibitem{0706.2363}
  M.~C.~N.~Cheng and E.~Verlinde,
  ``Dying Dyons Don't Count,''
  arXiv:0706.2363 [hep-th].

\bibitem{0802.0544}
  S.~Banerjee, A.~Sen and Y.~K.~Srivastava,
  ``Generalities of Quarter BPS Dyon 
Partition Function and Dyons of Torsion
  Two,''
  arXiv:0802.0544 [hep-th].

\bibitem{0802.1556}
  S.~Banerjee, A.~Sen and Y.~K.~Srivastava,
  ``Partition Functions of Torsion $>1$ Dyons in Heterotic
String Theory on $T^6$,''
  arXiv:0802.1556 [hep-th].

\bibitem{0803.1014}
  A.~Sen,
  ``N=8 Dyon Partition Function and Walls of Marginal Stability,''
  arXiv:0803.1014 [hep-th].

\bibitem{0803.2692}
  A.~Dabholkar, J.~Gomes and S.~Murthy,
  ``Counting all dyons in N =4 string theory,''
  arXiv:0803.2692 [hep-th].

\bibitem{9708062}
  A.~Gregori, E.~Kiritsis, C.~Kounnas, N.~A.~Obers, 
  P.~M.~Petropoulos and B.~Pioline,
  ``R**2 corrections and non-perturbative 
  dualities of N = 4 string ground
  states,''
  Nucl.\ Phys.\ B {\bf 510}, 423 (1998)
  [arXiv:hep-th/9708062].

\bibitem{9708130}
  E.~Kiritsis,
  ``Introduction to non-perturbative string theory,''
  arXiv:hep-th/9708130.

\bibitem{9504147}
  A.~Sen,
  ``Extremal black holes and elementary string states,''
  Mod.\ Phys.\ Lett.\  A {\bf 10}, 2081 (1995)
  [arXiv:hep-th/9504147].

\bibitem{0409148}
  A.~Dabholkar,
  ``Exact counting of black hole microstates,''
  Phys.\ Rev.\ Lett.\  {\bf 94}, 241301 (2005)
  [arXiv:hep-th/0409148].

\end{thebibliography}
\end{document}